\documentstyle[prb,aps]{revtex}
\begin{document}

\twocolumn[{
\draft
\widetext
\title{Time-dependent resonant tunneling via two discrete states}
\author{T.H. Stoof and Yu.V. Nazarov}
\address{
Department of Applied Physics and Delft Institute for
Microelectronics and 
Submicrontechnology,\\ Delft University of Technology, Lorentzweg 1,
2628 CJ Delft, The Netherlands}
\date{\today}
\maketitle
\vspace{-0.2in}
\mediumtext

\begin{abstract}
We theoretically investigate time-dependent resonant tunneling
via two discrete states in an experimentally relevant setup. Our
results show that the dc transport through the system can be
controled by applying external irradiation with a frequency which
matches the energy difference between the discrete states.
We predict resonant phenomena which should be easily observable
in experiments.
\end{abstract}

\bigskip
}]

\narrowtext

Time-dependent tunneling phenomena have received increasing
attention in recent years. In the early eighties,
B\"uttiker and Landauer studied the tunneling time needed for an
electron to traverse a potential barrier.\cite{butlan} More
recent theoretical work focused on the time-dependence of resonant
tunneling using an effective Schr\"odinger equation\cite{gurvitz}
and on a description of the time-dependent current through
mesoscopic structures in terms of non-equilibrium Green's
functions.\cite{wjm}
In addition, the considerable improvement in nanofabrication
techniques facilitated
some interesting experimental studies. Kouwenhoven {\it et al.}
measured the photon-assisted tunneling current through a single
quantum dot with an effectively continuous level spectrum due
to thermal smearing.\cite{leo}
Van der Vaart {\it et al.} studied the dc current through a double
dot system with well developed 0D states in each dot and clearly
resolved resonances between energy levels in both dots.\cite{nijs}
The sharp resonance features make it very tempting to
perform experiments with time-dependent fields.
The dc current through such a structure in the presence of
oscillating fields may be expected to display interesting new
phenomena, not observable in a single dot.

In this paper we use of the density matrix approach of
Ref.~\onlinecite{yuli}, in which
the resonant states are described by a time-dependent tunneling
Hamiltonian. Transitions between non-resonant states of the
system are taken into account through a master equation for the
density matrix elements. We calculate the photoresponse of the
system in several experimentally relevant limits and derive
an explicit expression for the shape of the resonant peaks in
the case of an external perturbation with arbirary amplitude.
Close to resonance the dc current is found to be very sensitive
to  the oscillating field. The satellite 
resonances induced by the external oscillating field can be
of the same order of magnitude as the main static resonance
with an even smaller width.

The system under consideration (Fig.~\ref{fig:system})
consists of two quantum dots $A$ and $B$ in series. The dots
are connected by tunnel junctions to two large reservoirs $L$ and
$R$, which are assumed to have continuous energy level spectra
and are filled up to their respective Fermi energies.
If we neglect all tunneling processes, a system of discrete
many-body states is formed in each dot. The best conditions for
transport occur when it costs no energy to transfer an electron
between the dots, {\it i.e.} the energy difference between 
a state with one extra electron in the left dot and a state with
one extra electron in the right dot is zero. In the experiment
in Ref.~\onlinecite{nijs} this energy difference could be tuned
by an external gate voltage. The current
through the system vs. gate voltage consists of a series of peaks
corresponding to the resonances between different discrete states.
There could be a variety of different transport processes
occuring in a resonance point as described in Ref.~\onlinecite{yuli}.
We concentrate on the simplest experimentally relevant case,
namely when
the resonance occurs between the ground states of both dots.
We assume that the bias voltage is much larger than the temperature
and the energy difference between the states in resonance. 
Consequently, electrons can only enter
the two-dot system from the left and leave it only to the right.
Transitions from the left and to the right lead are possible with
rates $\Gamma_{L}$ and $\Gamma_{R}$, respectively. Here and
throughout the paper units are used such that $\hbar=1$.
We will assume that it is impossible, due to the Coulomb blockade,
for an electron to tunnel into the system while another electron
is still present in either one of the dots.
We concentrate on two states only and disregard other states,
which is allowed in the neighborhood of the resonance.
The system can also be in a third state $|0\rangle$, when there is
no extra electron in either one of the dots. 
The energies of the resonant states, which both lie well between the
electrochemical potentials in the left and right lead, initially
differ by an amount $\epsilon_{0}$.
Under these conditions the transport
through the system depends only very weakly on the bias voltage, but
does depend strongly on the gate electrode via the energy difference
$\epsilon_{0}$.
We assume that a time-dependent oscillating signal is applied
to the gate electrode, so that the time-dependent energy difference
becomes: $\epsilon(t)=\epsilon_{0}+\tilde\epsilon \cos \omega t$,
where $\tilde\epsilon$ is the amplitude and $\omega$ the frequency
of the externally applied signal.

The dynamics of the resonant states, $|a\rangle$ and $|b\rangle$,
is governed by the time dependent tunneling Hamiltonian
${\cal H}(t)={\cal H}_{0}(t)+{\cal H}_{T}$, where ${\cal H}_{0}$
is given by
\begin{equation}
\label{hnul}
{\cal H}_{0}(t) = \frac{1}{2} \epsilon(t)
\left(\; |a\rangle\langle a| - |b\rangle\langle b| \;\right)
\end{equation}
and ${\cal H}_{T}$ describes the coupling between the dots that
introduces mixing between the eigenstates $|a\rangle$ and
$|b\rangle$ of the system:
\begin{equation}
\label{htunnel}
{\cal H}_{T} = T \left( \; |a\rangle\langle b| +
|b\rangle\langle a| \; \right) .
\end{equation}
The average current through the system is given by
\begin{equation}
\label{trace}
\langle I \rangle /e$=Tr$(\rho {\cal I}),
\end{equation}
where $\cal{I}$ is the current operator:
\begin{equation}
\label{curop}
{\cal I} = i T \left(\;|a\rangle\langle b| - |b\rangle\langle a| \;
\right),
\end{equation}
and $\rho$ is the density matrix for the two-level system.
We describe
transitions between different states in the density matrix
approach.\cite{yuli} The equations for the density
matrix elements read
\begin{mathletters}
\label{densmatel}
\begin{eqnarray}
\frac{\partial \rho_{a}}{\partial t} &=&
+ \Gamma_{L}\; \rho_{0} + iT \;(\rho_{ba}-\rho_{ab})\\
\frac{\partial \rho_{b}}{\partial t} &=&
-\Gamma_{R}\; \rho_{b} - iT \;(\rho_{ba}-\rho_{ab})\\
\frac{\partial \rho_{ab}}{\partial t} &=&
-\frac{1}{2} \Gamma_{R}\; \rho_{ab} + i\epsilon(t) \; \rho_{ab} +
iT \; (\rho_{b}-\rho_{a})\\
\frac{\partial \rho_{ba}}{\partial t} &=&
-\frac{1}{2} \Gamma_{R}\; \rho_{ba} - i\epsilon(t)\; \rho_{ba} -
iT \; (\rho_{b}-\rho_{a}),
\end{eqnarray}
\end{mathletters}
where $\rho_{a}$, $\rho_{b}$ and $\rho_{0}=1-\rho_{a}-\rho_{b}$
denote the probabilities for an electron to be in the left dot,
the right dot, or in neither dot, respectively, and
$\rho_{ab}=\rho_{ba}^{*}$ are the non-diagonal density matrix
elements. In these equations the terms proportional to $\Gamma_{L}$
and $\Gamma_{R}$ describe the transitions to and from
the reservoirs between the states $|0\rangle$ and $|a\rangle$
and the states $|b\rangle$ and $|0\rangle$, respectively.
All other terms follow from the Liouville equation:
$i d \rho /dt=[{\cal H},\rho]$. Note that the rates $\Gamma_{L}$
and $\Gamma_{R}$ do not enter the equations in a symmetric way.
$\Gamma_{R}$ describes the decay of the resonant states whereas
$\Gamma_{L}$ describes the build-up of these resonant states.

The relevant energy scales of the system are the transition rates
$\Gamma_{L}$ and $\Gamma_{R}$, the tunneling amplitude $T$, and
the frequency $\omega$ and amplitude $\tilde\epsilon$ of the
applied perturbation. There are three limiting cases for which we
can develop an analytical approach to the problem. They are
complementary and essentially cover all the interesting physics.

We will first consider the limiting case of a small perturbation
amplitude;
$\tilde\epsilon \ll \omega,T,\Gamma_{L,R}$.
Using the fact that $\rho_{0}=1-\rho_{a}-\rho_{b}$, we rewrite
Eqs.~(\ref{densmatel}) in matrix notation:
\begin{equation}
\label{densvect}
\frac{\partial \vec{\rho}}{\partial t} = \left(
\mathaccent 94\Gamma
+ \mathaccent 94T + \mathaccent 94\epsilon_{0} +
\mathaccent 94\epsilon \cos \omega t \right) \vec{\rho} + \vec{c},
\end{equation}
where $\vec{\rho}=(\rho_{a},\rho_{b},\rho_{ab},\rho_{ba})^{T}$,
$\vec{c}=(\Gamma_{L},0,0,0)^{T}$ and the matrices
$\mathaccent 94\Gamma$, $\mathaccent 94T$,
$\mathaccent 94\epsilon_{0}$ and
$\mathaccent 94\epsilon$ correspond to Eqs.~(\ref{densmatel}).
The stationary solution of these equations without irradiation is
\begin{equation}
\vec{\rho}_{0} =
-(\mathaccent 94\Gamma + \mathaccent 94T +
\mathaccent 94\epsilon_{0})^{-1} \vec{c}.
\label{rhonul}
\end{equation}
This determines the shape of the stationary resonant peaks
observed by Van der Vaart {\it et al.}:\cite{nijs}
\begin{equation}
I_{\mbox{\small{stat}}} = \frac{T^{2} \Gamma_{R}}{ T^{2}
(2+\Gamma_{R}/\Gamma_{L}) + \Gamma_{R}^{2}/4 + \epsilon_{0}^{2}}.
\label{statpeak}
\end{equation}
The first order correction to the stationary solution is
\begin{equation}
\vec{\rho}_{1}
= \vec{\rho}_{1}^{\;+} \exp(i \omega t) + \vec{\rho}_{1}^{\;-}
\exp(-i \omega t),
\label{rhoeen}
\end{equation}
with $\vec{\rho}_{1}^{\;\pm}$ the positive and negative frequency
part, respectively:
\begin{equation}
\vec{\rho}_{1}^{\;\pm} = -(\mathaccent 94\Gamma +
\mathaccent 94T + \mathaccent 94\epsilon_{0}
\mp i \omega \mathaccent 94I)^{-1}
\frac{\mathaccent 94\epsilon}{2} \vec{\rho}_{0},
\label{rhoplusmin}
\end{equation}
$\mathaccent 94I$ being the unit matrix.
This contribution contains only oscillatory terms, which average
out when calculating the dc current. We therefore also determine
the second order correction terms (proportional to
$\tilde\epsilon^{2}$) and obtain
\begin{equation}
\vec{\rho}_{2} = 
-(\mathaccent 94\Gamma + \mathaccent 94T +
\mathaccent 94\epsilon_{0})^{-1}
\frac{\mathaccent 94\epsilon}{2}
( \vec{\rho}_{1}^{\;+} + \vec{\rho}_{1}^{\;-} ).
\label{rhotwee}
\end{equation}
Using the density matrix elements $\vec{\rho}_{2}$
we may calculate the photoresponse of the system. This quantity
can be easily measured experimentally by slowly modulating the
irradiation amplitude.\cite{leo} 
In Fig.~\ref{fig:I-eps0-omega} a plot is given of the photoresponse
as a function of $\epsilon_{0}$ and $\omega$ for $\Gamma_{L}=
\Gamma_{R}=0.2 T$.
The figure clearly shows resonant
satellite peaks for $\omega$ and $\epsilon_{0}$ satisfying
$\omega^{2}=\epsilon_{0}^{2}+4T^{2}$, {\it i.e.} resonant modes
occur when the frequency of the applied perturbation matches the
renormalized energy difference $\sqrt{\epsilon_{0}^{2}+4T^{2}}$
of the two levels. For frequencies below $2 T$ there are no
satellite peaks because the energy $\hbar \omega$ of the photon
is smaller than the energy level spacing.
The evolution of a resonant satellite peak is shown in
Fig.~\ref{fig:evolution}, where a current peak for $\omega=3T$
has been plotted vs. $\epsilon_{0}$ for different values of
$\Gamma_{R}/T=\Gamma_{L}/T$.
We see that the peak can be seen even at moderately large values
of $\Gamma_{R}/T$, but the best resonance conditions occur when
$\Gamma_{R} \ll \mbox{max}(T,\epsilon_{0})$ and 
$\omega=\sqrt{\epsilon_{0}^{2}+4 T^{2}}$.

We have developed
a second approach which allows us to explore the satellite
peak at arbitrary values of irradiation amplitude $\tilde\epsilon$
under the conditions mentioned above. Substituting
$\vec{\rho} = \vec{\rho}_{0} + \vec{\rho}_{+}(t) \exp(i \omega t)
+ \vec{\rho}_{-}(t) \exp(-i \omega t)$ in Eq.~(\ref{densvect})
and neglecting terms proportional to $\exp(\pm 2 i \omega t)$ we
obtain:
\begin{mathletters}
\begin{eqnarray}
\label{dens2a}
\frac{\partial \vec{\rho}_{0}}{\partial t} & = &
\mathaccent 94\Gamma \vec{\rho}_{0} + (\mathaccent 94 \epsilon_{0}
+ \mathaccent 94T) \vec{\rho}_{0} +
\frac{\mathaccent 94 \epsilon}{2}
( \vec{\rho}_{+} + \vec{\rho}_{-} ) + \vec{c}, \\
\label{dens2b}
\frac{\partial \vec{\rho}_{+}}{\partial t} & = &
\mathaccent 94\Gamma \vec{\rho}_{+} + (\mathaccent 94 \epsilon_{0}
+ \mathaccent 94T -i \omega \mathaccent 94I) \vec{\rho}_{+} +
\frac{\mathaccent 94 \epsilon}{2} \vec{\rho}_{0},\\
\label{dens2c}
\frac{\partial \vec{\rho}_{-}}{\partial t} & = &
\mathaccent 94\Gamma \vec{\rho}_{-} + (\mathaccent 94 \epsilon_{0}
+ \mathaccent 94T + i \omega \mathaccent 94I) \vec{\rho}_{-} +
\frac{\mathaccent 94 \epsilon}{2} \vec{\rho}_{0}.
\end{eqnarray}
\end{mathletters}
Near the resonance point we can approximate the solution
$\vec{\rho}$ by an expansion in terms of the eigenvectors
of the matrix $\mathaccent 94 \epsilon_{0} + \mathaccent 94T$:
$\vec{\rho}_{0}=\alpha_{1} \vec{v}_{1}+\alpha_{2} \vec{v}_{2}$,
$\vec{\rho}_{+}=\alpha_{+} \vec{v}_{+}$ and 
$\vec{\rho}_{-}=\alpha_{-} \vec{v}_{-}$, where 
$\vec{v}_{1}$ and $\vec{v}_{2}$ are the eigenvectors with
eigenvalue $0$ and $\vec{v}_{\pm}$ those with eigenvalues
$\pm i \sqrt{\epsilon_{0}^{2}+4 T^{2}}$. We obtain a
set of four closed equations for the coefficients $\alpha_{1,2,+,-}$
by taking the inner product of Eq.~(\ref{dens2a}) with $\vec{v}_{1}$
and $\vec{v}_{2}$, of Eq.~(\ref{dens2b}) with $\vec{v}_{+}$,
and of Eq.~(\ref{dens2c}) with $\vec{v}_{-}$. Solving for the
stationary solution and calculating the current profile near
the resonance point $\epsilon_{r}=\sqrt{\omega^{2}-4 T^{2}}$
results in a Lorentzian line shape:
\begin{equation}
I/e = \frac{I_{\mbox{\small{max}}} w^{2}}{w^{2}+
(\epsilon_{0}-\epsilon_{r})^{2}},
\label{curpeak}
\end{equation}
with height
\begin{equation}
I_{\mbox{\small{max}}} = \frac{\tilde\epsilon^{2} \Gamma_{R}
(\alpha^{2}-4)}{\gamma(\gamma \Gamma_{R}^{2} + 
\beta \tilde\epsilon^{2})},
\label{height}
\end{equation}
and half-width at half maximum
\begin{equation}
w = \frac{\alpha}{2 \sqrt{\alpha^{2}-4}} \sqrt{ \Gamma_{R}^{2}
+ \frac{\beta}{\gamma} \tilde\epsilon^{2}},
\label{width}
\end{equation}
where $\alpha=\omega/T$, $\beta=\Gamma_{R}/\Gamma_{L}+2$,
and $\gamma=\alpha^{2}+\beta - 4$.

In the limit of small amplitude
the height of the current peak scales with the square of
$\tilde\epsilon$: $I_{\mbox{\small{max}}}=\tilde\epsilon^{2}
(\alpha^{2}-4)/\gamma^{2}\Gamma_{R}$, 
whereas the width remains constant:
$w=\frac{1}{2} \alpha \Gamma_{R} / \sqrt{\alpha^{2}-4}
\sim \Gamma_{R}$, consistent with the results presented in
Fig.~\ref{fig:evolution}. At $\alpha=2$ (corresponding to
$\epsilon_{0}=0$) the peak vanishes, as seen in
Fig.~{\ref{fig:I-eps0-omega}.
With further increase of $\tilde\epsilon$
the current saturates at a value of
$I_{\mbox{\small{sat}}}=\Gamma_{R}(\alpha^{2}-4)/\beta\gamma$
which is of the order
of the height of the stationary peak. This saturation occurs at
relatively small $\tilde\epsilon \sim \Gamma_{R}$. The width of the
peak increases with growing $\tilde\epsilon$. At $\tilde\epsilon
\gg \Gamma_{R}$ it is proportional to $\tilde\epsilon$.

Therefore we have shown that, under good resonance conditions,
the current is very sensitive to the external irradiation. A
relatively weak irradiation induces a big satellite peak that
has a much smaller width than the stationary one.

For small tunneling amplitudes $T$, provided
$\epsilon_{0} \gg \Gamma_{R}$, the height scales with $T^{2}$:
$I_{\mbox{\small{max}}}=T^{2}\tilde\epsilon^{2}/\Gamma_{R}
\omega^{2}$,
and the half-width reduces to $w=\frac{1}{2} \Gamma_{R}$. These
results agree with the expression for the photon-assisted
tunneling current derived below, where we consider our third
approach, in which the tunneling amplitude is small
compared to all other energy scales in the system;
$T \ll \tilde\epsilon,\omega,\Gamma_{L,R}$.

First we perform a transformation on the density matrix which
leaves the diagonal elements invariant and which changes the
non-diagonal elements as follows:
\begin{eqnarray}
\label{intpict}
\bar{\rho}_{ab} & = & \rho_{ab} \exp \left( -i
\int_{-\infty}^{t} d\tau \; \epsilon(\tau) \right),
\end{eqnarray}
This transformation eliminates the explicit time dependence in
Eqs.~(\ref{densmatel}c) and (\ref{densmatel}d) and introduces 
it in the transformed tunneling amplitude. The equations
for the non-diagonal density matrix elements now assume the form
\begin{equation}
\frac{\partial \bar{\rho}_{ab}}{\partial t} =
\left( i \epsilon_{0} -
\frac{1}{2}\Gamma_{R} \right) \bar{\rho}_{ab} +
i \bar{T}(t) \left(\bar{\rho}_{b}-\bar{\rho}_{a}
\right),
\label{rhoab}
\end{equation}
with time-dependent tunneling amplitude
\begin{equation}
\label{timedepT}
\bar{T}(t) = T \exp \left( i \int_{-\infty}^{t} d\tau \;
\epsilon(\tau) \right).
\end{equation}
The equations for $\bar{\rho}_{ba}$ are simply the complex conjugate
of Eqs.~(\ref{intpict}), (\ref{rhoab}) and (\ref{timedepT}).
For $\bar{\rho}_{ab,\omega}$ in the lowest non-vanishing order
in T we obtain:
\begin{equation}
\bar{\rho}_{ab,\omega} = \frac{i \bar{T}_{\omega}}{\frac{1}{2}
\Gamma_{R} + i (\omega - \epsilon_{0})}.
\label{rhoabomega}
\end{equation}
Expanding $\bar{T}(t)$ in a Fourier series;
\begin{equation}
\bar{T}(t) = T \sum_{n=-\infty}^{\infty}
J_{n}(\tilde\epsilon / \omega)
\exp \left( -i n \omega t\right),
\label{Treeks}
\end{equation}
and calculating the dc current, Eq.~(\ref{trace}), results in:
\begin{equation}
\langle I \rangle /e = T^{2} \Gamma_{R} \sum_{n=-\infty}^{\infty}
\frac{J_{n}^{2}(\tilde\epsilon / \omega)}{\frac{1}{4}
\Gamma_{R}^{2} + (n \omega -\epsilon_{0})^{2}},
\label{cur2}
\end{equation}
where $J_{n}$ is a $n$'th order Bessel function of the first kind.
 This equation for the current is similar to the expression
found by Tien and Gordon\cite{tiengordon} for the photon assisted
tunneling current through a superconducting tunnel junction.
Note, however, that in the Tien-Gordon case the current has been
considered as a function of bias voltage whereas in our case it
is a function of $\epsilon_{0}$, the energy shift of the discrete
levels. Analogously, the alternating field is not applied in
the bias direction but rather to the gate electrodes.

In Fig.~\ref{fig:I-eps0-eps} the current has been plotted as a
function of $\epsilon_{0}$ and $\tilde\epsilon$. The figure clearly
shows that the current is composed of a number of satellite peaks
each separated by the photon energy $\hbar \omega$. With increasing
amplitude $\tilde \epsilon$, the number of visible current peaks
increases. The peaks all have the same width $\Gamma_{R}$ and have
heights given by
$4 T^{2} J_{n}^{2}(\tilde\epsilon/\omega)/\Gamma_{R}$. 
In the limit of small amplitude the height of the $n=1$
satellite peak reduces to $I_{\mbox{\small{max}}}=
T^{2}\tilde\epsilon^{2}/ \Gamma_{R} \omega^{2}$, identical to
our earlier result.
Note that Eq.~(\ref{cur2}) for the current no longer contains
$\Gamma_{L}$. Because the tunnel rate from lead to dot is much
larger than the tunnel rate between the dots, the width of the
level is in this case determined by $\Gamma_{R}$ and $T$ only.

In conclusion, we have presented a complete theoretical picture
of the dc transport through a double quantum dot in the presence of
external harmonic irradiation. The photoresponse of the system
exhibits extra resonant peaks when the frequency of the external 
irradiation matches the energy difference between the discrete
states. At further increase of the irradiation intensity this
satellite peak becomes of the same order of magnitude as the
main peak, but preserves the much smaller width. At small tunneling
amplitudes and large irradiation amplitude extra satellite peaks
appear in a pattern similar to that obtained for a
Josephson junction by Tien and Gordon.\cite{tiengordon}

We acknowledge fruitful discussions with L.P. Kouwenhoven,
M. B\"uttiker, N.C. van der Vaart and G.E.W. Bauer.
This work is part of the research program of the "Stichting voor
Fundamenteel Onderzoek der Materie" (FOM), which is financially
supported by the "Nederlandse Organisatie voor Wetenschappelijk 
Onderzoek"
(NWO).

\begin{figure}[p]
\caption{Schematic picture of the system. Two quantum dots $A$
and $B$ are coupled to leads $L$ and $R$ via tunnel junctions.
Transitions are possible with rates $\Gamma_{L}$ and $\Gamma_{R}$.
The tunneling rate between the dots is $T$ and the energy difference
between the levels is denoted by $\epsilon(t)$.}
\label{fig:system}
\end{figure}

\begin{figure}[p]
\caption{Scaled photoresponse of the system as a function of the
energy difference $\epsilon_{0}/T$ between the levels and the
frequency $\omega/T$ of the applied signal. The plot was made with
$\Gamma_{L}=\Gamma_{R}=0.2T$.}
\label{fig:I-eps0-omega}
\end{figure}

\begin{figure}[p]
\caption{Evolution of a satellite peak for ratios
$\Gamma_{L}/T$= $\Gamma_{R}/T$= 0.2, 0.5, 1.0 and 3.0
respectively. The plots were made for a frequency
of $\omega=3T$.}
\label{fig:evolution}
\end{figure}

\begin{figure}[p]
\caption{Scaled current through the dots as a function of the
energy difference $\epsilon_{0}/T$ between the levels and the
amplitude $\tilde\epsilon/T$ of the applied signal. The plot was
made with $\Gamma_{R}=T$ and $\omega=10T$.}
\label{fig:I-eps0-eps}
\end{figure}

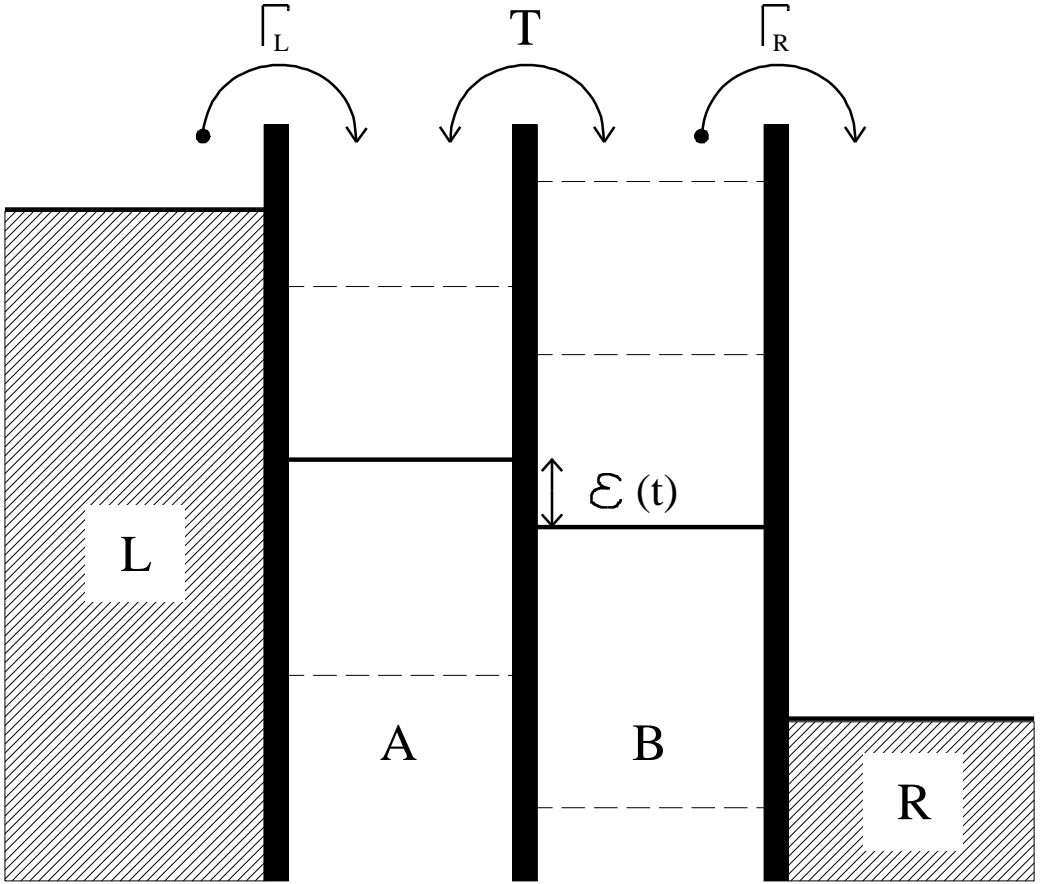
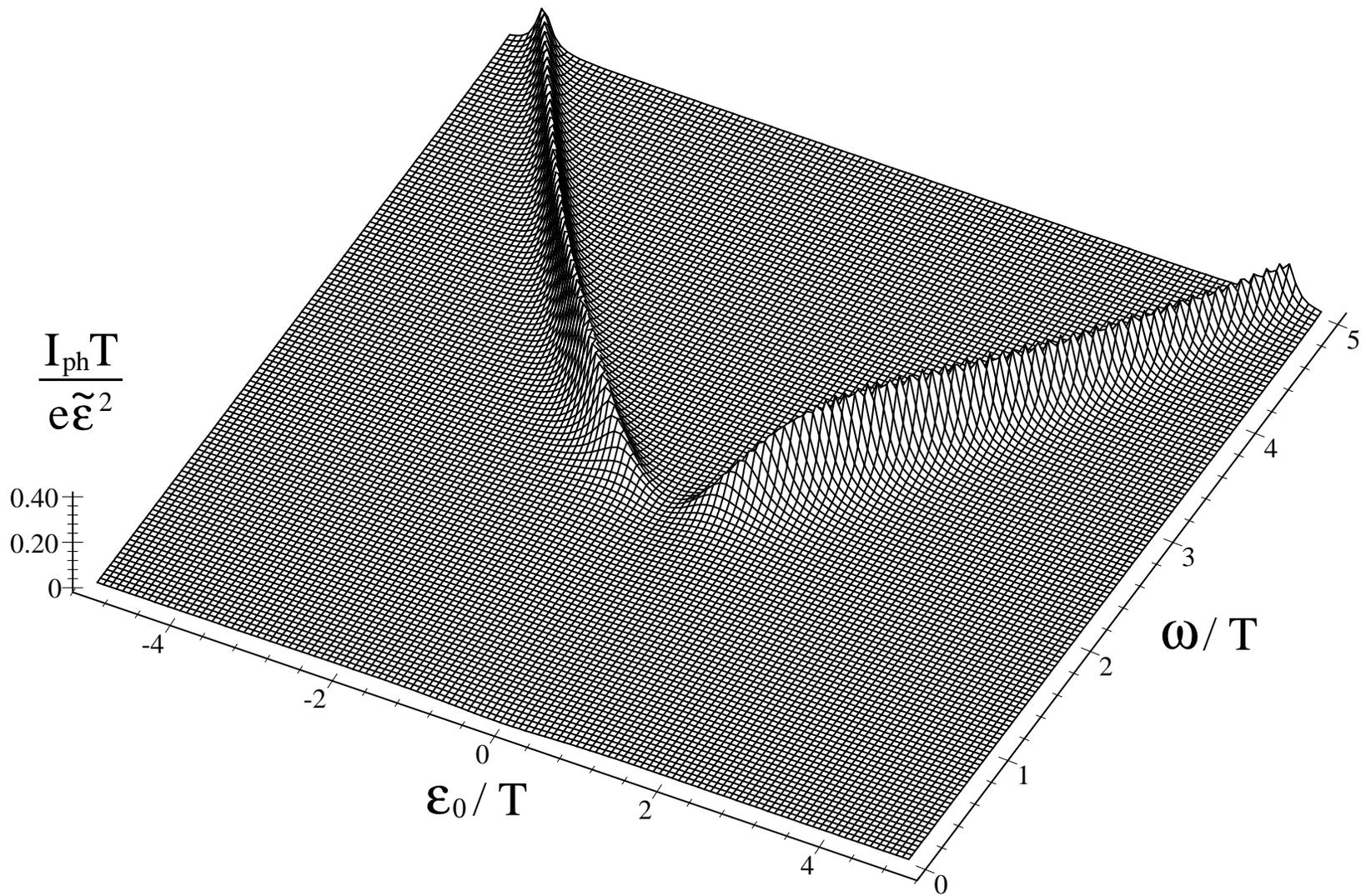
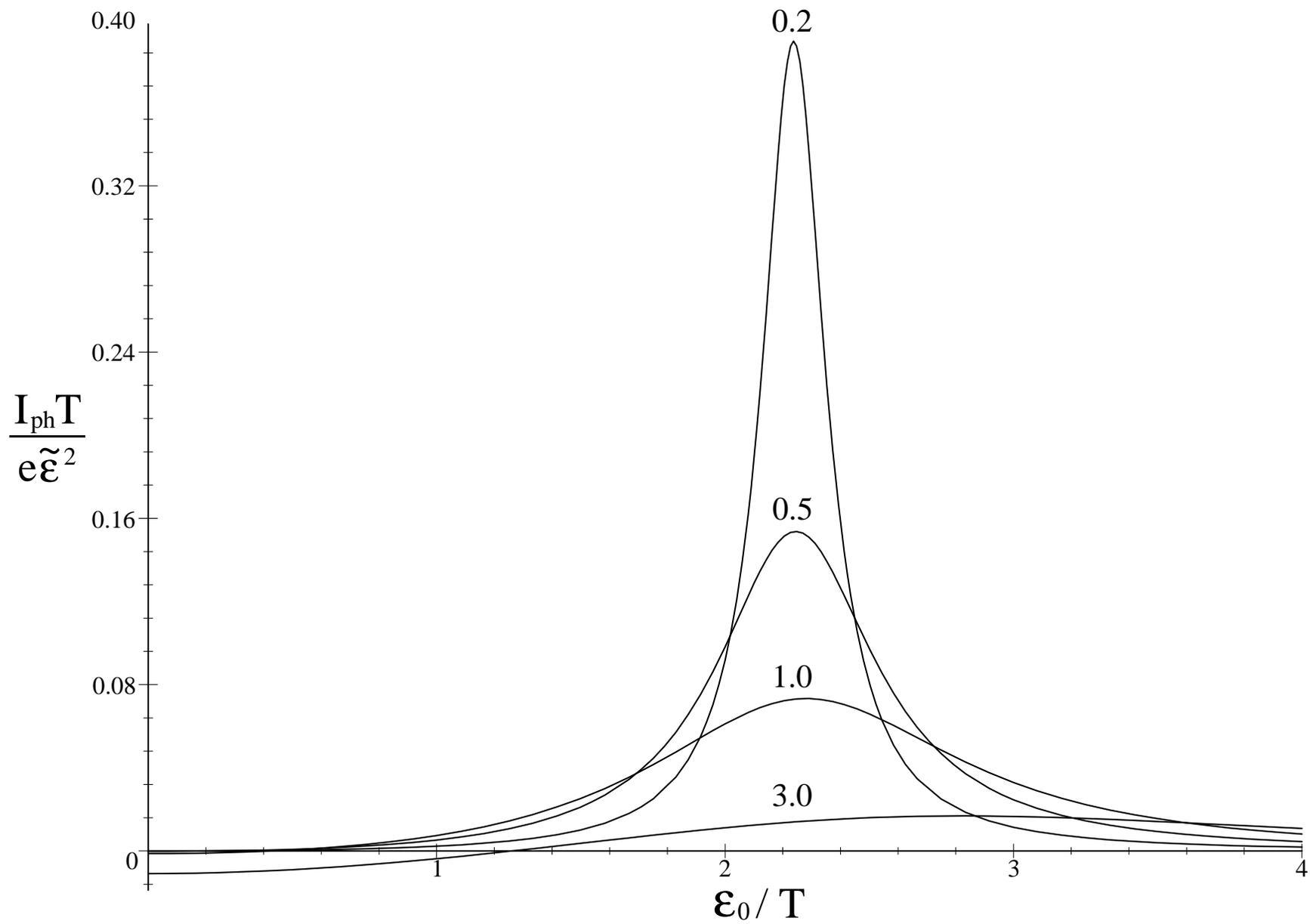
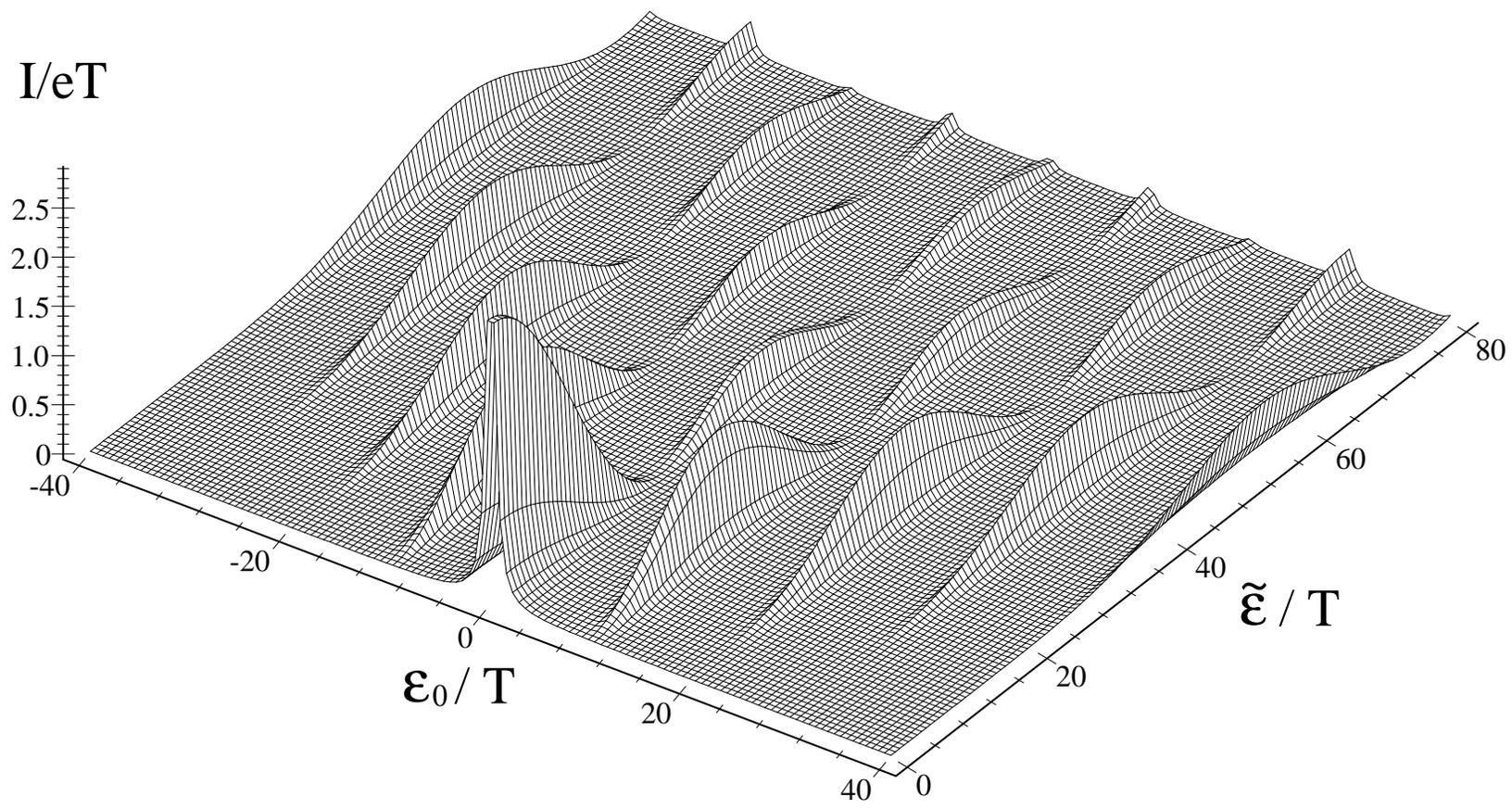
\end{document}